\input amstex
\documentstyle{amsppt}
\magnification=1200
\nopagenumbers

\centerline{\bf The Aharonov-Bohm effect in spectral asymptotics}

\centerline{\bf of the magnetic Schr\"odinger operator}

\centerline{G. Eskin and J. Ralston, UCLA}
\vskip.1in\centerline{\it In memory of Hans Duistermaat}
\vskip.2in \noindent {\bf Abstract}: We show that
in the absence of a magnetic field the spectrum
of the magnetic Schr\"odinger operator in an
annulus depends on the cosine of the flux
associated with the magnetic potential. This
result follows from an analysis of a singularity
in the \lq\lq wave trace" for this Schr\"odinger
operator, and hence shows that even in the
absence of a magnetic field the magnetic
potential can change the asymptotics of the
Schr\"odinger spectrum, i.e. the Aharonov-Bohm effect takes place.
 We also study the
Aharonov-Bohm effect for the magnetic
Schr\"odinger operator on a torus. \vskip.2in
\noindent {\bf \S 1. Introduction}.

\vskip.1in
 Let $\Omega$ be the exterior of a
bounded region in $\Bbb R^2$ with smooth
boundary, and let
$$H_{A,V}={1\over 2}(i\partial_{x_1}+A_1(x))^2+
{1\over 2}(i\partial_{x_2}+A_2(x))^2-V(x).$$
This is the Schr\"odinger operator for a particle
of mass 1 and charge -1 moving in $\Omega$ under
the influence of the magnetic potential
$A=(A_1,A_2)$ and the electric potential $V$. We
assume that
$$\partial_{x_2}A_1-\partial_{x_1}A_2=0 \hbox{ in }\Omega,\eqno{(1)}$$
i.e., the magnetic field vanishes in $\Omega$.
Given a simple, closed curve $\gamma$ in $\Omega$
encircling the complement of $\Omega$, we define
the magnetic flux by
$$\alpha_\gamma=\int_\gamma A(x)\cdot dx.$$
In view of (1) $\alpha_\gamma$ only depends on the orientation of $\gamma$.

In the seminal paper [AB] Aharonov and Bohm
showed that if $\alpha_\gamma\neq 0$ mod $2\pi$, then
one can detect the cosine of the magnetic flux in
the scattering of particles in this quantum
system, i.e. the magnetic potential has
a physical impact even when the magnetic field is zero in $\Omega$.
This is called the Aharonov-Bohm effect.
Aharonov and Bohm found this by computing the scattering cross-section
explicitly for $\Omega =\Bbb R^2\backslash\{0\}$,
when $A(x)=(-x_2/|x|^2,x_1/|x|^2)$ and $V(x)=0$.
They also proposed an experiment to demonstrate
this effect. However, the first generally accepted
experimental verification of the Aharonov-Bohm
(AB) effect was done many years later by Tonomura
et al. [T]. For further mathematical work on the
AB effect see [N], [W], [RY], [E2], [EIO].

\newpage
In [H] Helffer showed that $A(x)$ can influence
the spectrum of $H_{A,V}$ when the magnetic field
is zero in $\Omega$. In the semi-classical
setting with $V(x)\to\infty$, as $|x|\to\infty$,
and $\Omega=\{|x|>1\}$ he showed that the lowest
Dirichlet eigenvalue depended on the cosine of the magnetic
flux. Earlier related results on magnetic Schr\"odinger operators
are due to Lavine and O'Carroll ([L-C]).

In this paper we study the Schr\"odinger operator
in the domain $\Omega_R =\Omega\cap\{|x|<R\}$
with Dirichlet boundary conditions on $|x|=R$ and
$\partial \Omega$. We compute the singularity at
$t=3R\sqrt 3$ of the distribution trace of the
fundamental solution of the initial-boundary
value problem
$$u_{tt}+H_{A,V}u=0,\ u(x,0)=f(x),\ u_t(x,0)=0,\ u(x,t)=0\hbox{ when }x\in \partial\Omega_R.\eqno{(2)}$$
 This distribution trace is known as
the \lq\lq wave trace" for this problem, and it
is given by
$$\sum_{j=1}^\infty \cos(t\sqrt{\lambda_j}),$$
where $\{\lambda_j\}_{j=1}^\infty$ are the
Dirichlet eigenvalues of $H_{A,V}$ in $\Omega_R$.
Hence its singularities are determined by the
behavior of the $\lambda_j$ as $j\to\infty$.
These singularities are well-known to appear only
at the lengths of periodic broken ray paths in
$\Omega_R$. The singularity at $t=3R\sqrt 3$
comes from equilateral triangles in $\Omega_R$
with vertices on $|x|=R$. To compute this
singularity we need to know that $ 3R\sqrt 3$ is
isolated in the set of lengths of broken periodic
rays. To ensure that we assume that the
complement of $\Omega$, $\Omega^c$, is strictly
convex and contained in $\{|x|<1\}$ and $R\geq 8$
(see Remark I.1), but any assumption that makes
the length of the inscribed equilateral triangles
isolated in the lengths of periodic reflected ray
paths will suffice. The geometry that we have
chosen makes the singularity unchanged when one
changes the sign of $\alpha_\gamma$. Hence we
cannot recover more than the cosine of
$\alpha_\gamma$ from it (see Remark I.2).

A definitive computation of leading
singularities in wave traces was given by
Duistermaat and Guillemin in [DG] for manifolds
without boundary. For manifolds with boundary the
analogous computation has not been done in that
generality. To carry it out in here we have taken
this opportunity to present a different method of
computation that replaces Fourier integral
operators with superpositions of Gaussian beams (cf. [CRR] and Chapter 5 of [CR]). In \S 5 we briefly  discuss the computation of wave
trace singularities using the global theory of Fourier integral operators
(cf. [H], [D], [MF] and [E]).
Both approaches lead to the following:
\vskip.1 in
\noindent {\bf Theorem:} The  distribution
$$\sum_{j=1}^\infty \cos(t\sqrt{\lambda_j})$$
has an isolated singularity at $t=L=3R\sqrt 3$. The leading term in that singularity is
 the distribution
$$- 2^{-5/2}3^{1/4}R^{3/2}\cos(\int_\gamma A(x)\cdot
 dx)(t-L)_+^{-3/2},\eqno{(3)}.$$
Hence the wave trace determines the cosine of the magnetic flux.

\vskip.1in
In the final section of this paper we consider $H_{A,V}$ on (flat) 2-torus and obtain essentially
the same result: under a non-degeneracy assumption on the torus the singularities in the wave trace at times equal
to the lengths of curves in a homology basis determine the cosines of magnetic fluxes around those curves (see Theorem 5.1).

\noindent
{\bf Remark I.1:} The only fact from geometry
needed here -- and we only need it for circles -- is: a
ray and its reflections inside an ellipse are
all tangent to an ellipse confocal with the
boundary ellipse.
 So rays in $|x|\leq R$ tangent to a circle
$|x|=r>1$ will never enter $|x|<1$ after
reflection in $|x|=R$, while rays that enter
$|x|<1$ will always re-enter $|x|<1$ after
reflection in $|x|=R$. Since the boundary

\noindent curve $C$ is
convex, rays entering $|x|<1$ will leave
$|x|<1$ after at most one reflection. This gives
the following bounds on the length $L$ of
periodic ray paths that hit $C$. For rays that
close after entering $|x|<1$ $k$ times
$$2kR-2k<L<2kR+2k.$$
So periodic rays that enter $|x|<1$
more than three times have lengths are greater
than $8R-8$, and the equilateral triangles are
the (isolated) shortest periodic rays that never
enter $|x|<1$ (assuming $R>2$). So we need
$4R+4<3R\sqrt 3<6R-6$. That happens as soon as
$R\geq 8$ (picking the first whole number that
works). \vskip.1in \noindent{\bf Remark I.2:} If
$\Omega=\{|x|>1\}$ and $V\equiv 0$, the mapping
$u(x)\to u(-x)$ sends eigenfunctions of $H_{A,0}$
to eigenfunctions of $H_{-A,0}$ bijectively. Thus
the wave traces of these operators must be
identical. The leading singularity in the wave
trace at $t=3\sqrt 3 R$ does not depend on the
boundary of $\Omega$ or $V(x)$, hence it will be
unchanged when $A$ is replaced by $-A$ in these
cases, too. Therefore, one cannot distinguish
$\alpha_\gamma$ and $-\alpha_\gamma$ using the
leading singularity. The same ambiguity arises in
the results in [AB] and [H]. \vskip.2in
 \noindent {\bf \S 2.
Singularities of the Wave Trace}. \vskip.1in
 Let $E(x,y,t)$ denote the fundamental solution
for the initial-boundary value problem (2). The
 wave front set of
the distribution kernel of $E$ is contained in
the
 canonical relation for the
 bicharacteristic flow (see Melrose-Sj\"ostrand, [MS I, II]. For this problem this
 canonical relation is defined as follows:
 Let
 $\nu(x)$ denote the outer unit normal to
 $\partial \Omega_R$ at $x$. Given
 $(y_0,\eta_0)$ with $y_0\in \Omega_R$ and
 $|\eta_0|=1$, define $(x(s,y,\eta),\xi(s,y,\eta))=(y+s\eta,\eta)$
 until, at $s=s_1$, $y_1=x(s_1,\eta_0,y_0)\in \partial \Omega_R$. Then,
 if $\eta\cdot\nu(y_1)\neq 0$, continue
$(x(s,y_0,\eta_0),\xi(s,y_0,\eta_0))$ for $s>s_1$
as $(y_1+s\eta_1,\eta_1)$, where
$\eta_1=\eta_0-2(\nu(y_1)\cdot \eta_0)\nu(y_1)$.
Continue the bicharacteristic this way,
reflecting when $x(s,y_0,\eta_0)$ hits
$\partial\Omega_R$, as long as $x(s,y_0,\eta_0)$
does not intersect $\partial\Omega_R$
tangentially. At points of tangential
intersection one has to distinguish grazing and
gliding points. However, since we assume that the
boundary of $\Omega^c$ is strictly convex, points
of tangential intersection with $\partial \Omega$
are grazing points and bicharacteristics continue
unaffected by these intersections. When $y_0$ is
in the interior of $\Omega_R$, a bicharacteristic
with initial data $(y_0,\eta_0)$ will never
intersect $|x|=R$ tangentially. Hence, the wave
front set of the kernel of $E(\cdot,\cdot, t)$ is
the union over $y_0\in \Omega_R$ and $\eta_0\in
\Bbb S^1$ of the points
$$(x(t,y_0,\eta_0),\xi(t,y_0,\eta_0),y_0,-\eta_0),$$
where $(x(t,y_0,\eta_0),\xi(t,y_0,\eta_0))$ are the reflected bicharateristics described above.
 Strictly speaking, the wave front set is the closure of that set and includes a \lq\lq boundary
 wave front set" over $|x|=R$ (see [MS] for details).

Since $E(x,y,t)$ is a distribution in $t$
depending smoothly on $(x,y)\in \Omega_R\times
\Omega_R$, $\int_{\Omega_R}E(x,x,t)dx$ is
well-defined, and we have the following relation
$$T=_{def}\sum_{j=1}^\infty \cos(t\sqrt
\lambda_j)=\int_{\Omega_R} E(x,x,t)dx.$$

The
singular support of $T$ is contained in the set
of $t$ such that $(y_0,\eta_0,y_0,-\eta_0)\in
WF(E(x,y,t))$ for some $y_0\in
\overline{\Omega_R}$, [GM].
The choice of $\Omega$ and $R$ here implies that,
for $t$ in a sufficiently small neighborhood of
$3R\sqrt 3$, $(y_0,\eta_0,y_0,-\eta_0)\in
WF(E(x,y,t))$ only if the ray $x(s,y_0,\eta_0)$
traces an inscribed equilateral triangle.

To compute the singularities in the wave trace we
need a parametrix for the initial-boundary value
problem (2). Since this parametrix will differ
from $E(x,y,t)$ by an integral operator with a
smooth kernel, we can use it to compute
singularities. Since we are only interested in
singularities arising from inscribed equilateral
triangles, we only need a parametrix which
captures the singularities of $\int
E(x,y,t)f(y)dy$ when $WF(f)\subset \{y,\eta):
y\in \Omega_R, |y\cdot \eta^\perp|=R/2\}$, where
$(\eta_1,\eta_2)^\perp=(\eta_2,-\eta_1)$. These
singularities hit $\partial \Omega_R$
nontangentially, and hence this parametrix
construction can be done with reflection at the
boundary. This observation applies equally well
to constructions with Fourier integral operators
and the Gaussian beam superpositions used here.
\vskip.1in \noindent{\bf \S 3. The Gaussian beam
construction.} Here we will outline the
construction of a parametrix for (2), for
initial data with wave fronts projecting onto the
inscribed equilateral triangles. We will
continue to let $\eta$ have length one. The
Gaussian beam method allows one to do the
following (see [R] for more details): \vskip.1in \noindent i) For any ray,
$(x(t),t)=(z+t\eta,t)$, in space-time, one can
construct a function $\phi(x,t;z,\eta)$
satisfying: \vskip.1in (a) For any given integer
$N$, $(\phi_t)^2-|\phi_x|^2$ vanishes to order N
on $(x(t),t)$ and $\hbox{Im}\{\phi_{xx}\}$ is
positive definite on $(x(t),t)$. \vskip.1in (b)
$\phi(x,0;z,\eta)=x\cdot\eta+{i\over 2}|x-z|^2$
on $|x-z|<\delta$, and $\phi_t(x,0;z,\eta)=-1$.
\vskip.1in \noindent Moreover, if $\Gamma$ is a
curve with unit normal $\nu$ at $x(t_0)$ and
$\eta$ is not tangent to $\Gamma$, then one can
construct $\phi^r=\phi$ on $\Gamma$, satisfying
(a) for the reflected ray
$(x(t_0)+(t-t_0)\eta^r,t)$, where $\eta^r=\omega
-2(\nu\cdot \eta)\nu$. Reflection of beams is discussed in [R, \S 2.2].
 \vskip.1in \noindent ii)
Once $\phi$ has been constructed, for any given
integer $N$, one can solve the transport
equations
$$2\phi_t(a_0)_t-2\phi_x\cdot (a_0)_x+(2iA(x)\cdot
\phi_x +\phi_{tt}-\Delta \phi)a_0=0, \eqno{(4)}$$
$$2\phi_t(a_j)_t- 2\phi_x\cdot (a_j)_x+2iA(x)\cdot
\phi_x +\phi_{tt}-\Delta
\phi)a_j=-(\partial_t^2-(\partial_x+iA(x))^2)a_{j-1},\
j>0$$ to order $N$ on $(x(t),t)$,
 and impose
the initial conditions $a_0(0,x;z,\eta)=1$ and
$a_j((0,x;z,\eta)=0$ for $j>0$ on $|x-z|<\delta$.

\vskip.1in For the singularity computation we
need to know the leading amplitude $a_0$ on the
ray beginning at $z$ in direction $\eta$.

\vskip.1in
\noindent We define $a(x,t;z,\eta,r)$
to be the formal sum
$$a(x,t;z,\eta,r)=\sum_{j\geq 0}a_j(x,t;z,\eta)r^{-j}.\eqno{(5)}$$

\noindent As before one can reflect in a
plane curve $\Gamma$ which is transverse to the ray,
and we impose $a^r=-a$ on $\Gamma$ to satisfy
Dirichlet boundary conditions.

 \vskip.1in Using the preceding constructions we
can construct the operator
$$[V(t)f](x)={1\over 2}([V_+(t)f](x)+[V_-(t)f](x)),$$
where
$$[V_\pm(t)f](x)=\hskip4in $$
$$\sum_{k\geq 0}{1\over (2\pi)^3}\int_{\Bbb
R_+\times S^1\times \{|z|<R+\delta\}}e^{ir\phi^k(x,\pm
t;z,\eta)}a^k(x,\pm t;z,\eta,r)\hat
f(r\eta)r^2drd\eta dz.\eqno{(6)}$$
 Here,
$\phi^0$ is the phase function with
$\phi^0(x,0;z,\eta)=x\cdot\eta+{i\over
2}|x-z|^2$, and for $k>0$,
$$e^{ir\phi^k(x,t;z,\eta)}a^k(x,t;z,\eta,r)$$
is the (Dirichlet) reflection of
$e^{ir\phi^{k-1}(x,t;z,\eta)}a^{k-1}(x,t;z,\eta,r)$
in the circle $|x|=R$. Since Gaussian beams can
be constructed to for any finite ray segment, we
can assume that each term in (6) is defined on
$\{|x|\leq 2R\}$ when necessary. Note that in
this notation the variables $(z,\eta)$ in
$\phi^k$ remain the initial data {\bf at t=0} for
the ray where Im$\{\phi^k\}=0$. Note also that
the integration in $r$ in (6) is in the sense of
distributions.

For the parametrix construction we need
$V(0)f=f+Kf$ where $K$ is an operator with a
smooth kernel. From (6) we have
$$[V(0)f](x)={1\over (2\pi)^3}\int_{\Bbb
R_+\times S^1\times \{|z|<2R\}}e^{irx\cdot
\eta-r|x-z|^2/2} \hat f(r\eta)r^2drd\eta dz.$$ Since
$${1\over (2\pi)^3}\int_{\Bbb
R_+\times S^1\times \Bbb R_z^2}e^{irx\cdot
\eta-r|x-z|^2/2} \hat f(r\eta)r^2drd\eta dz
=f(x)$$ and $f$ is supported in $\{|x|<R\}$, it
follows that omitting the contribution from
$\{|z|>R+\delta \}$ in (6) only adds an operator
with a smooth kernel.

 \vskip.1in To compute
singularities of the wave trace we need to make
the kernels of the operators $V_\pm(t)$ explicit.
The distribution kernels of these operators are
sums of terms of the form
$$S(t)=\int_{\Bbb
R_+\times S^1\times \Bbb R_z^2}e^{ir\phi(x,t;z,\eta)-ir\eta\cdot
y}a(x,t;z,\eta,r) r^2drd\eta
dz,\eqno{(7)}$$

\noindent As was stated earlier, these operators
are smooth in $(x,y)$, and we can compute their
traces by
  integrating these kernels
 over the diagonal $y=x$. Thus the (distribution)
trace of $V(t)$ is a sum of terms of the form
$$Tr(\phi,a)=\int_{D\times \Bbb
R_+\times S^1\times \Bbb
R_z^2}e^{ir\phi(x,t;z,\omega)-ir\eta\cdot
x}a(x,t;z,\eta,r) r^2drd\eta dzdx.\eqno{(8)}$$

\noindent We want to compute the
singularity in $t$ of this trace at $t=L=3R\sqrt 3$, and we
only need to consider $t$ in $|t-L|<\delta$,
where $\delta$ is small enough that $\{t:
|t-L|<\delta\}$ contains no other lengths of
periodic rays in the disk $|x|<R$.

\vskip.1in \noindent {\bf \S 4. Calculation of
the singularity at $t=L=3\sqrt 3 R$} \vskip.1in
For $\eta=(\eta_1,\eta_2)$ with $|\eta|=1$ define
$\eta^\perp=(\eta_2,-\eta_1)$, the \lq\lq right
hand" normal. To compute the singularity at $t=L$
we only need the parametrix restricted to
$R/2-\epsilon<|z\cdot \eta^\perp|<R/2 +\epsilon$ for
any fixed positive $\epsilon$. Since the broken ray
$x(t,z,\eta)$ is initially of the form
$x=z+t\eta$, $\eta^\perp\cdot z>0$ corresponds to
rays going counterclockwise around $z=0$, and
$\eta^\perp\cdot z<0$ corresponds to rays going
clockwise around $z=0$.

In the preceding section we concluded that the singularity in the wave trace at $t=L$
could be calculated from a sum of integrals of the form
$${1\over 2}\sum_\pm\int_0^\infty r^2dr\int_{\Cal S^1}d\eta\left(\int
a_0 (x,\pm t,z,\eta)e^{ir(\phi(x,\pm t,z,\eta)-x\cdot
\eta)}dxdz\right).\eqno{(9)}$$ The
integral in $r$ is to be taken in distribution
sense. Until the end of this section we will consider (9) in the case that the phase $\phi$ is the beam phase
resulting from reflecting the bicharacteristic
with initial data $(x,\xi)=(z,\eta)$ three times
in $|x|=R$. The amplitudes $a_0(x,t,z,\eta)$
are determined by the transport equation (4). The contributions to the singularity
from the $+$ and $-$ terms in (9) are complex conjugates of each other, and from here one
we only consider the \lq\lq$+$" term.

 We assume that that $a_0$
vanishes when $|z\cdot \eta^\perp|$ is not
close to $R/2$. Note that we can assume that
$\phi(x,t,z,\eta)$ is defined for all $(x,z,t)$
when $|z\cdot\eta^\perp|$ is sufficiently close
to $R/2$.

The main step in isolating
the singularity is an application of the method
of stationary phase to (9). For that we introduce
the change of coordinates
$$x=u+v\eta+w\eta^\perp,\ z=v\eta +w\eta^\perp,
\ u\in \Bbb R^2,\ v,w\in \Bbb R.$$ Our
objective is the elimination of the integral in
$(u,w)$ by stationary phase. To see when the
phase is real and stationary in these variables
note that \vskip.1in \noindent i) the phase is
real only when $x=x(t,z,\eta)$, \vskip.1in

\noindent ii) the derivative of the phase with
respect to $u$ at $x=x(t,z,\eta)$ is
$$\phi_x-\eta=\xi(t,z,\eta)-\eta,$$
which vanishes precisely when three reflections
have made $\xi$ return to its initial value. That
implies $|z\cdot \eta^\perp|=R/2$. Since the
reflected ray will return to $z$ when $t=L$ and
it is propagating in the direction $\eta$,
$x(t,z,\eta)=z+(t-L)\eta$.
Hence $u=(t-L)\eta$
and $|w|=R/2$ on the stationary set in $u$.
 The derivative of the phase
with respect to $w$ at $x=x(t,z,\eta)$ is
$$\eta^\perp\cdot \phi_x+\eta^\perp\cdot
\phi_z-\eta\cdot \eta^\perp$$ which vanishes,
since
$\phi_z(x(t,z,\eta),t,z,\eta)=\phi_z(x(0,z,\eta),0,z,\eta)
=\partial_z(x\cdot \eta +i|x-z|^2/2)|_{x=z}=0$.
Thus we will need to do the stationary phase
computation at $(u,w)=((t-L)\eta, \pm R/2)$.

\vskip.1in Calculation of asymptotics by stationary phase requires the
computation of the determinant of the Hessian of the phase, and here this
computation is rather long. We have found it
useful to consider the phase and the
bicharacteristics defined for all $\eta\neq 0$ by
homogeneity. That makes the Jacobian matrix
$$F(t)=\left(\matrix {\partial x\over \partial
z}(t,z,\eta)&{\partial x\over \partial
\eta}(t,z,\eta)\\
{\partial \xi\over \partial z}(t,z,\eta)&{\partial
\xi\over \partial \eta}(t,z,\eta)\endmatrix
\right)=_{def}\left(\matrix
a&b\\c&d\endmatrix\right)$$ symplectic. Using
$\phi_x(x(t,z,\eta),t,z,\eta)=\xi(t,z,\eta)$ and
$\phi_z(x(t,z,\eta),t,z,\eta)=0$ and setting
$M=\phi_{xx}(x(t,z,\eta),t,z,\eta)$, one computes
directly that at $x=x(t,z,\eta)$
$$H=_{def}\left(\matrix \phi_{xx}&\phi_{xz} \\ \phi_{zx}&\phi_{zz}\endmatrix \right)
=\left( \matrix M &c-Ma \\ c^t-a^tM &a^tMa-a^tc\endmatrix \right).$$

\vskip.1in \noindent Letting $O_\eta$ be the
matrix with columns $\eta$ and $\eta^\perp$, one
sees that the Hessian of the phase in (9) with
respect to the variables $(u,v,w)$ is $B^tHB$
where
$$B=\left(\matrix I&O_\eta\\0 & O_\eta\endmatrix
\right).$$ However, we need the Hessian with
respect to $(u,w)$. We will see that
$\left(\matrix \eta\\ \eta \endmatrix\right)$ is
a null vector for $H$, and we have
$B\left(\matrix
0\\0\\1\\0\endmatrix\right)=\left(\matrix \eta\\
\eta \endmatrix\right)$. Moreover, letting
$P_\eta$ denote the orthogonal projection of
$\Bbb R^2$ onto $\langle \eta \rangle$, one
computes
$$B^t\left(\matrix 0&0\\0&P_\eta
\endmatrix\right)B=\left(\matrix 0&0&0&0\\ 0&0&0&0\\ 0&0&1&0\\0&0&0&0\endmatrix\right).$$
Hence,
$$\hbox{ det }\left(\matrix
\phi_{u_1u_1}&\phi_{u_1u_2}&0&\phi_{u_1w}\\
\phi_{u_2u_1}&\phi_{u_2u_2}&0&\phi_{u_2w}\\0&0&1&0\\
\phi_{wu_1}&\phi_{wu_2}&0&\phi_{ww}\endmatrix
\right) =\hbox{ det }\left( \matrix M &c-Ma \\
c^t-a^tM &a^tMa-a^tc+P_\eta\endmatrix \right).\eqno{(10)}$$

  To proceed with this computation we
need to know $F(t)$. The computation begins with
the formulas for $x(t,z,\eta)$ and
$\xi(t,z,\eta)$ after three reflections:
$$x(t,z,\eta)=w{\xi^\perp\over |\xi|}+(t+{z\cdot\eta\over |\eta|}-6\sqrt{R^2-w^2}){\xi\over |\xi|}
$$
and, setting $\eta=|\eta|(\cos \theta,\sin
\theta)$,
$$\xi(t,z,\eta)= |\eta|(\cos(\theta+ \pi-6\sin^{-1}{w\over R}),\sin(\theta+\pi-6\sin^{-1}{w\over R})).$$

\vskip.1in \noindent One checks that
$\partial_zw={\eta^\perp\over |\eta|}\hbox{ and
}\partial_\eta w=-(z\cdot \eta){\eta^\perp\over
|\eta|^3}$, and this implies that the Jacobian
${\partial\xi\over \partial z}$ at $w=\pm R/2$ is
${4\sqrt 3\over R}|\eta|P_{\eta^\perp}$. So
$c={4\sqrt 3\over R}|\eta|P_{\eta^\perp}$. Using
$\partial_\eta\theta=-\eta^\perp/|\eta|^2$, one
finds that at $w=\pm R/2$
$${\partial \xi\over \partial \eta}=P_\eta+P_\eta^\perp -{4\sqrt 3\over R}
{z\cdot\eta\over \eta}P_{\eta^\perp}=I-{4\sqrt 3\over R}{z\cdot\eta\over |\eta|}P_{\eta^\perp}$$
So $d=I-{4\sqrt 3\over R}vP_{\eta^\perp}$.

\vskip.1in
 The computations of the derivatives of
$x(t,z,\eta)$ are longer, but they are simplified
by the observation that $|\xi(t,z,\eta)|=|\eta|$.
At $w=\pm R/2$ one has
$${\partial x\over \partial z}=P_{\eta^\perp}\mp 2\sqrt 3{\eta\over |\eta|}\langle {\eta^\perp\over |\eta|},\cdot\rangle+
{\eta\over |\eta|}\langle
{\eta\over|\eta|}\pm 2\sqrt 3{\eta^\perp\over
|\eta|},\cdot \rangle+ (t-L+{z\cdot \eta \over
|\eta|}){4\sqrt 3\over R}P_{\eta^\perp} $$
$$=I+(t-L+{z\cdot \eta\over
|\eta|}){4\sqrt3\over R}P_{\eta^\perp}.$$ So $a=I+(t-L+v){4\sqrt
3\over R}P_{\eta^\perp}$. \vskip.1in To compute
${\partial x\over \partial \eta}$ at $w=\pm R/2$
one uses
$$({\xi\over |\xi|})_\eta = {1\over
|\eta|}(1-{4\sqrt 3\over R}{z\cdot \eta\over
|\eta|})P_{\eta^\perp}$$ at $w=\pm R/2$, and the less obvious result
that
$$({\xi^\perp\over
|\xi|})_\eta=(-1+{4\sqrt 3\over R}{z\cdot \eta\over |\eta|}){\eta\over |\eta|^2}\langle {\eta^\perp\over |\eta|},\cdot\rangle.$$
 Combining those with $\partial_\eta v=(z\cdot
\eta^\perp){\eta^\perp\over |\eta|^3}=\pm {R\over
2|\eta|^2}\eta^\perp$, one has

$${\partial x\over \partial
\eta}={\eta^\perp\over |\eta|}\langle
-(z\cdot\eta){\eta^\perp\over |\eta|^3},
\cdot\rangle \pm {R\over 2}(-1+{4\sqrt 3\over
R}{z\cdot \eta\over |\eta|}){\eta\over
|\eta|^2}\langle {\eta^\perp\over
|\eta|},\cdot\rangle$$ $$+ {\eta\over
|\eta|}\langle (\pm{R\over
2|\eta|^2}\eta^\perp\mp 2\sqrt 3(z\cdot
\eta){\eta^\perp\over |\eta|^3},\cdot
\rangle+(t-L+{z\cdot \eta\over |\eta|})(1-{4\sqrt
3\over R}{z\cdot\eta\over
|\eta|})P_{\eta^\perp})|\eta|^{-1}$$
$$={(t-L+{z\cdot \eta\over |\eta|})\over |\eta|}
(1-{4\sqrt 3\over R}{z\cdot \eta\over |\eta|})P_{\eta^\perp}
-{(z\cdot \eta)\over |\eta|^2}P_{\eta^\perp}.$$ Thus, when $(5\sqrt 3)/2-v<t<(7\sqrt 3)/2-v$,
$$F(t)=\left(\matrix I+(t-L+v){4\sqrt 3\over R}P_{\eta^\perp}&{(t-L)
\over
|\eta|}(1-{4\sqrt 3\over R}v )
 P_{\eta^\perp}-{4\sqrt 3\over R}{v^2\over |\eta|}P_{\eta^\perp}\\ {4\sqrt 3\over R}|\eta|P_{\eta^\perp}& I-{4\sqrt 3\over R}vP_{\eta^\perp}
\endmatrix\right).\eqno{(11)}$$

From this point onward we will assume that $|\eta|=1$, i.e. $\eta=(\cos \theta,\sin \theta)$. Note that this implies $|\xi(t,z,\eta)|\equiv 1$.

\vskip.1in

Now we can resume the computation of the Hessian.
First we compute the determinant of the Hessian.
For this the only facts that we need from the computation
of the symplectic matrix $F(t)$ -- it is a good
check on the computation to verify that it {\it
is} symplectic -- are that $a$, $b$, $c$ and $d$
commute with $P_\eta$ with $aP_\eta=dP_\eta
=P_\eta$ and $bP_\eta=bP_\eta=0$. We will also
eventually use the exact form of $c$. Note that
since $F(t)$ is symplectic $a^tc$ and $d^tb$ are
symmetric and $a^td-c^tb=I$.

\vskip.1in
 Returning
to (10) we have
$$\left( \matrix M &c-Ma \\
c^t-a^tM &a^tMa-a^tc+P_\eta\endmatrix \right)\left(\matrix I&a\\0&I\endmatrix \right)
=\left(\matrix M&c\\c^t-a^tM&P_\eta\endmatrix \right)\hbox{ and}$$
$$\left(\matrix I&0\\a^t&I\endmatrix\right)\left(\matrix M&c\\c^t-a^tM&P_\eta\endmatrix
\right)=\left(\matrix M&c\\c^t&a^tc
+P_\eta\endmatrix \right).$$ Since $M=(c+id)(a+ib)^{-1}$ (cf. [CRR]),
$$\left(\matrix M&c\\c^t&a^tc +P_\eta\endmatrix
\right)\left(\matrix a+ib &0\\0&I\endmatrix\right)=\left(\matrix c+id &c\\c^ta+ic^tb&a^tc+P_\eta\endmatrix \right)$$
$$\left(\matrix-a^t&I\\I&0\endmatrix\right)\left(\matrix c+id &c\\c^ta+ic^tb&a^tc+P_\eta\endmatrix
\right)=\left(\matrix
i(c^tb-a^td)&P_\eta\\c+id&c\endmatrix
\right)=\left(\matrix
-iI&P_\eta\\c+id&c\endmatrix\right).$$ Finally
$$\left(\matrix -ic+d&I\\I&0\endmatrix\right)\left(\matrix
-iI&P_\eta\\c+id&c\endmatrix\right)=\left(\matrix
0&P_\eta+c\\-iI&P_\eta\endmatrix\right).$$ From
the preceding, using the exact form of $c$, one
can read off the determinant of the Hessian of
the phase (at $u=(t-L)\eta$, $w=\pm R/2$). It is
$$(-1)({4\sqrt3\over R})\hbox{ det}((a+ib)^{-1}).\eqno{(12)}$$


\vskip.1in
At this point it is convenient to calculate the amplitude
$a_0$. Note that $\phi_t(a_0)_t-\phi_x\cdot (a_0)_x=-{d\over dt}a_0(x(t,z,\eta),t,z,\eta)$. Hence (4) implies that, after three
reflections,
$$a_0(x(t,z,\eta),t,z,\eta)=(-1)^3e^{i\int_0^tA(x(s))\dot
x(s)ds}e^{(\int_0^t[\phi_{tt}-\Delta
\phi](x(s),s)ds)/2}.\eqno{(13)}$$ Note that $|\phi_x|+\phi_t$ vanishes to second order when $x=x(t,z,\eta)$ and thus
$\phi_{tt}+\phi_{tx}\cdot\dot x = 0$ and $\phi_x=\xi(t,z,\eta)$ when $x=x(t,z,\eta)$.
Differentiating $|\phi_x|+\phi_t=0$ with respect to $x$ and using $\phi_{tt}=-\phi_{tx}\cdot\dot x$, we have
$\phi_{tt}-\Delta \phi= \xi\cdot M\xi-\hbox{trace}(M),$
when $x=x(t,z,\eta)$.

Differentiating $\dot x=\xi/|\xi|$ with respect to $z$ and $\eta$ and restricting to $|\eta|=1$ one sees that
 $\dot a +i\dot b=(I-P_\xi)(c+id)$. Hence, using $M=(c+id)(a+ib)^{-1}$, we see that, when $x(t,z,\eta)$ is not a reflection point,
 $${d\over dt}(\hbox{ log det} (a+ib))=\hbox{trace}((\dot a +i\dot b)(a+ib)^{-1})=\hbox{trace}((I-P_\xi)M )= \Delta \phi-\phi_{tt}.\eqno{(14)}$$
 At reflection points $a+ib$ jumps to $(1-2P_\nu)(a+ib)$, where $\nu$ is normal to the boundary. Thus $\hbox{det}(a+ib)$ is multiplied by $-1$. Note that, since the imaginary part of $M$ is positive definite and the trace of $(I-P_\xi)M$ equals the trace of $(I-P_\xi)M(I-P_\xi)$, (14) shows that the argument of $\hbox{det}(a+ib)$ is strictly increasing away from reflection points. Thus we can make the argument of $(\hbox{det}(a+ib))^{1/2}$ increasing by defining it to be 1 when $t=0$, to be multiplied by $i$ at each reflection point, and to be continuous between reflection points.
  With this definition of $(\hbox{det}(a+ib))^{1/2}$, we can conclude that after three reflections
$$a_0(x(t,z,\eta),t,z,\eta)=i(\hbox{det}(a+ib))^{-1/2}e^{i\int_0^tA(x(s))\dot
x(s)ds}.\eqno{(15)}$$

 We have $\int_0^LA(x(s))\dot
x(s)ds=\alpha_\gamma$, where $\gamma$ is the equilateral triangle traced by $x(s,z,\eta)$ with $z=v\eta+ (R/2)\eta^\perp$ or $z=v\eta- (R/2)\eta^\perp$. Since the magnetic field vanishes in $\Omega$, $\alpha_\gamma$ is independent of $v$ and $\eta$, and its value when $z=v\eta+(R/2)\eta^\perp$ is the negative of its value when $z=v\eta- (R/2)\eta^\perp$.

 Now we can evaluate the integral in
$(u,w)$ asymptotically by the method of
stationary phase. The standard
form of the stationary phase lemma, ([H\"or], Theorem 7.7.5), gives
the following: if $f(y)$ is a smooth function
such that Im$\{f\}\geq 0$, $f_y(y_0)=0$ and the
Hessian $f_{yy}(y_0)$ is nonsingular, then for
$a$ smooth with support in a sufficiently small
neighborhood of $y_0$, one has the asymptotic
expansion
$$\int_{\Bbb R^n}e^{irf(y)}a(y)dy=\left({2\pi\over
r}\right)^{n/2}\sum_{j=0}^\infty c_jr^{-j},$$
 and the leading coefficient is given by
$$c_0=e^{irf(y_0)}a(y_0)(\hbox{det}(-if_{yy}(y_0))^{-1/2}.\eqno{(16)}$$
Here the square root of the determinant in $(\hbox{det}(-if_{yy}(y_0))^{-1/2}$ is the analytic continuation to symmetric matrices with nonnegative real part of the positive
square root for positive definite matrices, see [H\"or, Theorem 7.7.5]. \vskip.1in
In our case we will use stationary phase to
eliminate the integrations in $u$ and $w$ in (9)
-- recall that $z=v\eta +w\eta^\perp$ and
$x=u+v\eta +w\eta^\perp$. The stationary point
$y_0$ in (13) is either $(u,w)=((t-L)\eta,
R/2)$ or $(u,w)=((t-L)\eta, -R/2)$ . Since
$$\phi(x(t,z,\eta),t,z,\eta)=\phi(x(0,z,\eta),0,z,\eta)=z\cdot
\eta,$$ and we have
$$f(y_0)=\phi(x(t,z,\eta)
,t,z,\eta)-x(t,z,\eta)\cdot\eta
$$
evaluated at $(u,w)=((t-L)\eta, R/2)$ or
$(u,w)=((t-L)\eta, -R/2)$, it follows that
$f(y_0)=-(t-L)$. The domain of integration in $(u,v,w,\eta)$ is
$$\{(u,v,w,\eta): |\eta|=1,\
|u+v\eta+w\eta^\perp|\leq R \hbox{ and }
\sqrt{w^2+v^2}<R+\delta\}.\eqno{(17)}$$ We consider (9) as an iterated integral with the integrations in $(u,w)$ done first. After we use the stationary phase lemma in those integrations, the resulting integrand is evaluated at $(u,w)=((t-L)\eta, \pm R/2)$, and, since we can assume that $|t-L|$ is smaller than $\delta$, the domain of integration in $(v,\eta)$ becomes
$$D=_{def}[-{\sqrt 3\over 2}R-(t-L),{\sqrt 3\over 2}R-(t-L)]\times S^1.$$ The stationary phase argument needs to be modified when $v$ is near $\pm\sqrt 3R/2$. There, since the integration in $(u,w)$ should not cross $|x|=R$, the stationary phase lemma does not apply. However, there is a simple remedy for this. Let $\rho= |u+v\eta+w\eta^\perp|$. On the sphere $\rho=R$ we can introduce coordinates $(\theta_1,\theta_2,\theta_3)$, functions of $(u,w)$ depending on $v$ as a parameter, near the points $(u,v,w)=((t-L)\eta,\pm \sqrt 3R/2,\pm R/2)$. Next using smooth cutoffs one can write the trace integral as the sum of an integral over a region when $\rho<R-\delta$, where the stationary phase argument applies as given earlier, and a region where $R-2\delta <\rho<R$. In the second region, near the points where the phase is stationary, one writes the integral in the variables $(\theta_1,\theta_2,\theta_3,v,\eta)$, and applies stationary phase in $(\theta_1,\theta_2,\theta_3)$. The stationary set will be the image in these coordinates of $(u,w)=((t-L)\eta,\pm R/2)$ and it will depend on $v$. Likewise, letting $Q$ denote the hessian in $(u,w)$ of the phase at the stationary points, the hessian at the stationary points will now be $J^tQJ$, where $J$ is the jacobian matrix of $(u,w)$ with respect to $(\theta_1,\theta_2,\theta_3)$. Since the $\theta$ variables are tangential, one can use the stationary phase expansion {\it uniformly} in $v$. The leading term will be an integral over the stationary set. On that set $(\hbox{det } Q)^{-1/2}$ will be replaced by
 $(\hbox{det }J^tQJ)^{-1/2}=|\hbox{det } J|^{-1}(\hbox{det} Q)^{-1/2}$. However, the new factor $|\hbox{det }J|^{-1}$ is canceled by the jacobian in the volume form (we have $dudw= |\hbox{det } J|d\theta_1d\theta_2d\theta_3$). Hence, the stationary phase expansion holds uniformly up to $v=\pm \sqrt 3R/2$. The result is that (12), (15) and (16) give uniformly for $(v,\eta)\in D$

$$\int_{D(v,\eta)}a_0 (x,t,z,\eta)e^{ir(\phi(x,t,z,\eta)-x\cdot
\eta)}dudw
=\pm {c(R)\over r^{3/2}}K(t)e^{-ir(t-L)}+ O({1\over r^{5/2}}),\eqno{(18)}$$
where $D(v,\eta)=\{(u,w): |u+v\eta+w\eta^\perp|\leq R\}$,
 and $c(R)=(2\pi )^{3/2}({R\over 4\sqrt 3})^{1/2}e^{3\pi i/4}$. The choice of sign $\pm$ is determined by (15) and (16): it is +1 when the square roots of det$(a+ib)$ implicit in (15) and (16) agree and -1 when they do not. The factor $$K(t)=\exp(i\int_0^tA(x^+(s))\cdot\dot
x^+(s)ds) +\exp(i\int_0^tA(x^-(s))\cdot
\dot x^-(s)ds)$$
 arises from adding the contributions from stationary points with $w=-R/2$ and $w=R/2$. The path $x^-(s)$ with $w=-R/2$ goes clockwise around the origin, and the path $x^+(s)$ with $w=R/2$ is counterclockwise. Hence $K(L)=2\cos(\int_\gamma A(x)\cdot
dx)$.
\vskip.1in
To compute the singularity we need the distribution calculation
  $$ \int_0^\infty e^{-i(t-L)r}r^{1/2}dr$$ $$={e^{-3\pi i/4}\Gamma(3/2)\over (t-L-i0)^{3/2}}= e^{-3\pi i/4}\Gamma(3/2)(t-L)_+^{-3/2}+e^{3\pi i/4}\Gamma(3/2)(t-L)_-^{-3/2}, \eqno{(19)}$$
where the homogeneous distributions $(s)^{-3/2}_\pm$ are defined by integration by parts and vanish on functions supported in $\mp s>0$. Note that the contribution to the trace from $V_-(t)$ is the complex conjugate of the contribution from $V_+(t)$. Hence, integrating over $(v,\eta, r)$, and adding the contributions from $V_(t)$ and $V_+(t)$ gives the leading singularity in the trace at $t=L$ as
$$\pm 2^{-5/2}R^{3/2}3^{1/4}\cos(\int_\gamma A(x)\cdot
 dx)(t-L)_+^{-3/2}. \eqno{(20)}$$
 The computation up to this point has not determined the choice of sign ($\pm$) in (20). That will be done in Remark 4.1, and there is an alternative derivation in \S 5. However, since the choice of sign in (20) does not depend on $A$, (20) is sufficient to conclude that the trace determines the cosine of the magnetic flux.

 \vskip.1in
 The final step in this argument is showing that (20) really is the leading term in the singularity. We have not discussed the contributions of the beams with phases $\phi^j$ in (6) for $j\neq 3$. However, those phases are never stationary near the periodic orbits, and give smooth contributions to the trace by the \lq\lq non-stationary phase" argument. Note that we can apply that argument up to $|x|=R$ by using the coordinates $(\theta_1,\theta_2,\theta_3)$ as before.

\vskip.2in
\noindent {\bf Remark 4.1} The sign \lq\lq $\pm$" in the leading singularity is actually \lq\lq -". To verify that we need to determine the signs of $(\hbox{det} (a+ib))^{1/2}$ in both the stationary phase computation and the amplitude computation.
\vskip.1in
 We begin with the stationary phase calculation. The matrix on the right in (10) can be rewritten as
$$\tilde H= \left( \matrix M &c-Ma \\
c^t-a^tM &a^tMa-a^tc+P_\eta\endmatrix \right)=\left(\matrix (c+id)(a+ib)^{-1}&-i(a+ib)^{-1}\\-i(a^t+ib^t)^{-1}&i(a+ib)^{-1}a +P_\eta\endmatrix\right)$$
This is a consequence of $F(t)$ being a symplectic matrix. Then, using (11) with $t=L$, one sees that $\tilde H$ has the invariant subspaces $V_1=\langle (\eta,\eta),(\eta,-\eta)\rangle$ and $V_2=\langle (\eta^\perp,\eta^\perp),(\eta^\perp,-\eta^\perp)\rangle$. The product of the eigenvalues of $\tilde H$ from eigenvectors in $V_1$ is $i$ (the eigenvalues are $1/2+(1\pm\sqrt 3/2)i$) and the product of the eigenvalues from eigenvectors in $V_2$ is $iC(A+iB)^{-1}$ where $A=\eta^\perp\cdot a\eta^\perp$, $B=\eta^\perp\cdot b\eta^\perp$ and $C=\eta^\perp\cdot c\eta^\perp$. Since all the eigenvalues have non-negative imaginary parts, this makes
 $$(\hbox{det}(-i\tilde H))^{-1/2}= {\sqrt{A+iB}\over \sqrt C}e^{i\pi/4}= {1\over 2}R^{1/2}3^{-1/4}e^{i\pi/4}\sqrt{A+iB},$$
in the stationary phase formula,
where $\sqrt{A+iB}$ is in the lower half-plane. That $\sqrt{A+iB}$ here is in Im$\{z\}<0$ is the point of the calculation, note that $A+iB=\hbox{det} (a+ib)$ at $t=L$.
\vskip.1in
To calculate $(\hbox{det} (a+ib))^{-1/2}$ in the amplitude we need to consider the entire ray path tracing an equilateral triangle beginning at $z=(z\cdot \eta)\eta \pm (R/2)\eta^\perp$ when $t=0$ and returning to that point when $t=L$. Without loss of generality we will assume that $z=(z\cdot \eta)\eta + (R/2)\eta^\perp$. Recall that $a(t)+ib(t)={\partial x\over \partial z}(t,z,\eta)+i{\partial x\over \partial\eta}(t,z,\eta)$. As we observed in the calculation of the amplitude $a_0$, $\hbox{det} (a+ib))$ is multiplied by $-1$ at each reflection. Geometric optics, following the reflection rule in Remark I.1, shows that, after the first reflection at $(x,t)=((\sqrt 3 R/2)\eta+(R/2)\eta^\perp,\sqrt 3 R/2-z\cdot \eta )$, there is exactly one \lq\lq focal point" where $\hbox{det}({\partial x\over \partial z})=0$ on each side of the triangle. Moreover, the homogeneity of $x(t,z,\eta)$ in $\eta$ of degree zero, implies that ${\partial x\over \partial \eta}\eta\equiv 0$. That implies that the real part of $\hbox{det} (a(t)+ib(t))$ changes sign from negative to positive at the points where $\hbox{det}({\partial x\over \partial z})=0$. Since the argument of $\hbox{det} (a(t)+ib(t))$ is increasing, this makes it possible to track the its change as $t$ goes from 0 to $L$: the total change when the path reaches the third focal point is $2\pi+2\pi +3\pi/2$. Since the argument of $(\hbox{det} (a(0)+ib(0)))^{1/2}$ was chosen to be zero, this means that at the third focal point, its argument will be $3\pi/4$ and $(\hbox{det} (a(L)+ib(L)))^{1/2}$ will be in the upper half plane. Thus, the choices of $(\hbox{det} (a(L)+ib(L))^{1/2}$ in the stationary phase computation and the amplitude computations have opposite signs, and the sign of the leading singularity in (18) is \lq\lq -".
\vskip.2in
\noindent {\bf Remark 4.2} We used triangular periodic orbits here because it was easy to give conditions that would make their lengths isolated in the set of lengths of periodic orbits (Remark 4.2). However, it is easy to extend the trace formulas for periodic orbits which are regular N-gons. These would give the same results when one can show that their lengths are isolated in the lengths of periodic orbits.
\vskip.1in
For a regular inscribed N-gon the length of a side is $h_N= 2R\sin {\pi\over N}$, and its total length is $L_N=Nh_N$. For the N-gon the entries in the first column of the jacobian from (11) become
$${\partial x\over \partial z}(t,z,\eta)=I+(t+v-L_N){4N\over h_N}P_{\eta^\perp}\qquad \hbox{and}\qquad
{\partial \xi\over \partial z}(z,\eta)={4N\over h_N}P_{\eta^\perp}.$$
 One can use either the analysis in Remark 4.1 or the Fourier integral approach in \S 5 to show that the only change this makes in the leading singularity is replacing the factor of $(\sqrt 3 R)({R\over 4\sqrt 3})^{1/2}$, which arose from integration in $v$ and $(\hbox{ det }{\partial \xi\over \partial z})^{-1/2}$ from the stationary phase, by $(h_N)({h_N\over 4N})^{1/2}$ and replacing the initial $-1$ in (20) -- note that $\pm$ is - by Remark 4.1 -- by $(i)^{N-1}$, since there is one focal point on each side. If one combines that with (19) and (20), the result is that the leading singularity in the trace is
$$(-1)^{(N-1)/2}C(N,\alpha_\gamma)(t-L_N)_+^{-3/2}\hbox{ for N odd, and }\eqno{(21a)}$$
$$(-1)^{N/2-1}C(N,\alpha_\gamma)(t-L)_-^{-3/2}\hbox{ for N even,}\eqno{(21b)}$$
$$\hbox{where}\quad C(N,\alpha_\gamma)=2^{-5/2}h_N^{3/2}N^{-1/2}\cos( \alpha_\gamma)={1\over 2}N^{-1/2}(R\sin{\pi\over N})^{3/2}\cos(\alpha_\gamma).$$

\vskip.2in
\noindent {\bf \S 5. A Fourier Integral Approach.} This problem provides an opportunity for direct comparison of Gaussian beam superpositions and Fourier integral operators. In this section we describe the computation of the singularities in the wave trace using global Fourier integral operators as in [H], [D], [MF] and [E]. This method requires a detailed description
of the singularities in the projection of bicharacteristics to
$x$-space, but in a simple situations like ours one can arrive at the formula for the leading singularity quickly. There are analytical arguments needed to justify that computation, and we will sketch them. Both methods make essential use of the computations of ${\partial x\over \partial z}$ and ${\partial \xi\over \partial z}$ in (11).
\vskip.1in
Let $E(t)$ be the fundamental solution for the boundary value problem (3). We will construct a parametrix for $E(t)$, micro-localized near the periodic rays, as a global Fourier integral operator. For $f$ supported in $\Omega_R$ let
$$[W(t)f](x)=[W_+(t)f](x)+[W_-(t)f](x)$$ $$={1\over 2(2\pi)^2}\int_{\Bbb R^2}(W_+(x,t,\eta)+W_-(x,t,\eta))\hat f(\eta)d\eta,$$
where
$$[W_\pm(0)f](x)= {1\over 2(2\pi)^2}\int_{\Bbb R^2}e^{ix\cdot \eta}\hat f(\eta)d\eta ={1\over 2}f(x).$$
Since the analysis of $W_+(t)$ and $W_-(t)$ is the same, we will work with $W_+(t)$ from here on.
\vskip.1in
The kernel $W_+(x,t,\eta)$ is given by $\exp(-it|\eta|+ix\cdot \eta)$ plus terms arising from reflection in $|x|=R$, Of course, the
phase and amplitude develop singularities, and in a neighborhood of those the form of $W_+(t)$ is more complicated, involving integrals over auxiliary variables. The Schwartz kernel of $W_+(t)$ is given by
$$\int_{\Bbb R^2}W_+ (x,t,\eta) e^{-iy\cdot \eta}d\eta.$$
This is a distribution in $t$ depending smoothly on $(x,y)$. Hence, the distribution trace of $W_+(t)$ is given by
$$ \int_{\Omega_R}(\int_{\Bbb R^2}e^{-ix\cdot\eta}W_+(x,t,\eta)d\eta)dx. \eqno{(22)}$$
\vskip.1in
Denote the reflected bicharacteristics with initial data $(x(0),\xi(0))=(z,\eta)$ by $(x(t,z,\eta),\xi(t,z,\eta))$ as in \S 2. We will write $\eta=|\eta|\hat \eta$ with $\hat \eta=(\cos \theta,\sin \theta)$ and $\hat \eta^\perp=(\sin\theta,-\cos\theta)$. Note that, since $x(t,z,\eta)$ is homogeneous of degree zero in $\eta$, we have $x(t,z,\eta)=x(t,z,\hat\eta)$. In what follows $\hat \eta$ will be treated as a parameter; all estimates will be uniform in $\hat\eta\in \Bbb S^1$. We will use the coordinates $(v,w)$ in $x$-space, where $x=v\hat \eta + w\hat \eta^\perp$, and the coordinates $(\tilde v,\tilde w)$ in $z$-space, where $z=\tilde v \hat \eta +\tilde w\hat \eta^\perp$. Since only periodic ray paths contribute to the singularities of the wave trace, we only need to consider $(\tilde v,\tilde w)$ with $|\tilde w-R/2|<\delta$ or $|\tilde w+R/2|<\delta$. Since the analysis is identical in both cases, we will only consider $|\tilde w-R/2|<\delta$. We are only interested in $t$ close to $L$. For convenience of notation we will use
$(x(\tilde v,\tilde w),\xi(\tilde v,\tilde w))=_{def}(x(L,\tilde v \hat \eta +\tilde w\hat \eta^\perp,\hat \eta), \xi(L,\tilde v \hat \eta +\tilde w\hat \eta^\perp,\hat \eta)).$
\vskip.1in
We will use the formulas for bicharacteristics after three reflections that were used to derive (11). From those formulas one sees that when $t=L$ the Jacobian $\partial (v,w)/ \partial (\tilde v,\tilde w)$ vanishes on the set
$\tilde\Sigma$
 where $\tilde v=(35/6)\sqrt{R^2-\tilde w^2}-L$. We define $\Sigma$ to be the image under the mapping $x=x(\tilde v,\tilde w)$ of the intersection of $\tilde \Sigma$ with $|\tilde w-R/2|<\delta$. The set $\Sigma$ is usually called the \lq\lq caustic set" for the bicharacteristics.
\vskip.1in
Let $\chi_0(z,\hat\eta),\chi_\pm(z,\hat\eta)$  be $C^\infty$  functions in
$\tilde U=\{|\tilde w- {R\over 2}|<\delta,|\tilde v|<\sqrt{R^2-{\tilde w}^2}\}$
equal to zero near $|\tilde w-{R\over 2}|=\delta$  and such that
$\chi_0(z,\hat\eta)=0$  for
$|\tilde v-\tilde v(\tilde w)|>2\epsilon,
\ \chi_+(z,\hat\eta)=0$
for
$\tilde v-\tilde v(\tilde w)<\epsilon,$ and $
 \chi_-(z,\hat\eta)=0$
for
$\tilde v-\tilde v(\tilde w)>-\epsilon$,
where
$\tilde v=\tilde v(\tilde w)$  is the equation of
$\tilde\Sigma,$, and $ \epsilon$  is fixed.
We assume also that $\chi_0+\chi_++\chi_-=1$   for $|\tilde w-{R\over 2}|
<{\delta\over 2}$.  Denote by $\tilde G_\pm$  the supports of
$\chi_0,\chi_\pm$,  respectively,   and let $G_\pm$  be the images of
$\tilde G_\pm$  under the mapping  $x=x(\tilde v,\tilde w)$.
Denote by $V_0(x,t,\eta)e^{-iz\eta}, V_\pm(x,t,\eta)e^{-iz\eta}$  the distribution
kernels corresponding  to the initial conditions
${1\over 2(2\pi)^2} \chi_0(z,\hat\eta)e^{i(x-z)\cdot\eta}$,
${1\over 2(2\pi)^2} \chi_\pm(z,\hat\eta)e^{i(x-z)\cdot\eta},$
respectively.
Note  that the difference
$W_+(x,t,\eta)-(V_0(x,t,\eta)+V_+(x,t,\eta)+V_-(x,t,\eta))$
does not contribute to the singularity near $t=L$.
\vskip.1in
It follows from [MF], [E,\S 66]  that $V_\pm(x,t,\eta)$
has the following  form  on
$G_\pm: V_\pm(x,t,\eta)=V_\pm^0(x,t,\eta)(1+R^\pm(x,t,\eta))$,
where
$$
V_\pm^0(x,t,\eta)={(-1)^3\over 8\pi^2}
\chi_\pm(z^\pm(x,t,\hat\eta),\hat\eta)|\hbox{det}
{\partial x^\pm\over\partial z}|^{-1/2}
\exp(i[{\pi\over 4}\sigma^\pm +\alpha(t)+\phi^\pm(x,t,\eta)]), \eqno{(23)}
$$
and
$R^\pm\approx \sum_{k\geq 1}r_k^\pm(x,t,\hat \eta)|\eta|^{-k}$  is
an asymptotic series in $|\eta|$.
Here
$\phi^\pm(x,t,\eta)=z^\pm(x,t,\hat\eta)\cdot\hat \eta$, where $z=z^\pm(x,t,\eta)$  is
the inverse function  to
$x=x(t,z,\eta)$  in $\tilde G_\pm$, and ${\partial x^\pm\over\partial z}=
{\partial x\over \partial z}(t,z^\pm(x,t,\hat\eta),\hat \eta)$.
The piecewise constant function $\sigma^\pm$ in (23) is the sum
of the \lq\lq phase shifts" at the focal points on the ray paths used
to define $\phi^\pm$. The sum of these phase shifts along the curve
$x(t,z,\eta),\ 0\leq t\leq L$ is called \lq\lq Maslov index" of this curve (see [MF, \S 1.7] or [E, \S 66]). The computation of the phase shifts at the focal points here can be done as in [E, 66.46-48], and the result is that the contribution to  $\sigma$ is -2 for
each focal point that $x(t,\tilde v\eta + R/2\eta^\perp,\eta)$
has passed through up to time $t$. This makes $\sigma^+=\sigma^- -2.$  The function $\alpha(t)=\int_0^t A(x(s,z,\eta))\cdot \dot x(s,z,\eta)ds$, and the factor $(-1)^3$ comes from the three reflections of
a  ray on $0\leq t\leq L$. Note that $V_\pm$
decay rapidly in $|\eta|$  outside $G_\pm$, respectively.
\vskip.1in
We denote the leading term of $\int_{\Omega_R}(V_1+V_2)e^{-ix\cdot\eta}dx$
by $I(t,\eta)=I_++I_-$,  where $I_\pm(t,\eta)
=e^{-i|\eta|(t-L)}\int_{G_\pm}V_\pm^0(x,L,\eta)e^{-ix\cdot\eta}dx$.
The phase in $I_\pm(t,\eta)$  is $\Phi_\pm(x,L,\eta)$

\noindent $=\phi_\pm(x,L,\eta)-x\cdot\eta$. The phase functions $\phi^\pm(x,t,\eta)$
  satisfy
$\phi_t^\pm+|\phi_x|^2=0$, and we have
$$\phi_x^\pm(x(t,z,\eta),t,\eta)=\xi(t,z,\eta),\
\phi_\eta^\pm(x(t,z,\eta),t,\eta)=z.\eqno(24)$$
Since  $|\phi_x^\pm|=|\eta|$  we have
$\phi_t^\pm=-|\eta|.$  Therefore
$\phi^\pm(x,t,\eta)=\phi^\pm(x,L,\eta)-|\eta|(t-L).$
The critical points  of $\Phi^\pm(x,L,\eta)$  are solutions  of
$\phi_x^\pm(x,L,\eta)-\eta=0,\ \phi_\eta^\pm(x,L,\eta)-x=0$.
It follows from (24) that $\xi(L,z,\eta)=\eta$  and $z=x(L,z,\eta)$.
In the geometry here this means that
the periodic orbit is an equilateral triangle inscribed
in $|x|\leq R$, and $L=3R\sqrt 3$.  Since any point of this triangle is a critical point, we need
to use the stationary phase expansion in the  transversal variable $w$.
\vskip.1in
Note that $z^\pm(x,L,\eta)=x=v\hat \eta+{R\over 2}\hat\eta^\perp,\ x\in G_\pm$.
Hence
$\Phi^\pm(v\hat\eta +{R\over 2}\hat \eta^\perp,L,\eta)
=\phi^\pm(v\hat\eta+{R\over 2}\hat \eta^\perp,L,\eta)-x\cdot\eta=0$.
Also
$\Phi_w^\pm(v\hat\eta +{R\over 2}\hat \eta^\perp,L,\eta)
=\phi_x^\pm(v\hat\eta+{R\over 2}\hat \eta^\perp,L,\eta)\cdot \hat\eta^\perp=0$,
since $\phi_x^\pm -\eta=0$  and $\eta\cdot\hat\eta^\perp=0$.  Compute now
$\Phi_{ww}^\pm(v\hat\eta +{R\over 2}\hat \eta^\perp,L,\eta)
=\hat\eta^\perp\cdot \phi_{xx}^\pm(v\hat\eta+{R\over 2}\hat \eta^\perp,L,\eta)\hat\eta^\perp$.
Differentiating
$\phi_x^\pm(x,L,\eta)=\xi(L,z^\pm(x,L,\eta),\eta)$  in $x$  we get
$\phi_{xx}^\pm={\partial\xi\over\partial z}({\partial x\over\partial z})^{-1}$
at $x=v\hat\eta+{R\over 2}\hat\eta^\perp,\ x\in G_\pm$.
It follows from (11)  that $\Phi_{ww}^\pm(v\hat\eta+{R\over 2}\hat\eta^\perp,L,\eta)=
{4\sqrt 3\over R}(1+v{4\sqrt 3\over R})^{-1}$.  Note that  $\Phi_{ww}^\pm>0$
when $v>-{R\over 4\sqrt 3}$  and   $\Phi^\pm_{ww}<0$
when $v<-{R\over 4\sqrt 3}$.
\vskip.1in
At this point we have the data needed in the stationary phase formula, but we need to consider the behavior of the amplitude that comes from (23). Since $\hbox{det}{\partial x\over \partial z}(L,z,\eta)=1+  v{4\sqrt 3\over R}$,
the factor $|\hbox{det}{\partial x\over \partial z}|^{-1/2}$ in the amplitude is canceled by part of the factor $|\Phi_{ w  w}^\pm|^{-1/2}$
in the stationary phase formula.  Hence the stationary phase expansion in $w$
has the leading terms
$${(-1)^3\over 8\pi^2}
({2\pi\over |\eta|})^{1/2}({R\over 4\sqrt 3})^{1/2}
\chi_-( v,{R\over 2},\hat\eta)
\exp(i[(L-t)|\eta|+{\pi\over 4} \sigma^- +\alpha(L)-\pi/4]),$$
$$ \hbox{ for }
v<-R/(4\sqrt 3);$$
$${(-1)^3\over 8\pi^2}
({2\pi\over |\eta|})^{1/2}({R\over 4\sqrt 3})^{1/2}
\chi_+(v,{R\over 2},\hat\eta)
\exp(i[(L-t)|\eta|+{\pi\over 4} \sigma^+ +\alpha(L)+\pi/4]),$$
$$ \hbox{ for } v>-R/(4\sqrt 3),$$
where $\sigma^-$ and $\sigma^+$ are the values of $\sigma$ before and after crossing the focal point at
$v=-R/(4\sqrt 3)$. Since $\sigma^-=\ -4$ and $\sigma_+=\ -6$,
the two formulas above can be combined to give the leading term in the integrand
 in (23) after integration in $w$
 $$ {2(\chi_++\chi_-)\over 8\pi^2}\cos(\alpha(L))({2\pi\over |\eta|})^{1/2}({R\over 4\sqrt 3})^{1/2}\exp(i[(L-t)|\eta|-\pi/4])\eqno(25)$$
 Here we have included the contributions from both
$w=R/2$ and $w=-R/2$ which have $\alpha(L)$ with opposite signs.
\vskip.1in
Now we will find the contribution of $\int_{\Omega_R}V_0(x,t,\eta)e^{-ix\cdot\eta}dx$.
The caustic set
$\Sigma$  is a fold type singularity  (cf. [D] and [E, Example 66.1]).
Therefore $V_0(x,t,\eta)$  is given  by an integral representation (cf. (66.53) in [E],
see also [L])
$$
V_0(x,t,\eta)={|\eta|^{1\over 2}e^{i(L-t)|\eta|}\over(2\pi)^{1\over 2}}
\int_{-\infty}^\infty a(v,\xi_2,|\eta|)e^{i|\eta|(S(v,\xi_2,L)+w\xi_2)}d\xi_2.
\eqno(26)
$$
Computing the stationary points in (26)  for $x\in G_-\cap \{d(x,\Sigma)<\epsilon\}$  we see that
the stationary points are given by $S_{\xi_2}(v,p^-(v,w),L)+w=0$  and the phase  is
$S(v,p^-,L)+wp^-=\phi^-(x,t,\eta)$,  where  $\phi^-(x,t,\eta)$ is the same as in (23).
The amplitude $a(v,\xi_2,|\eta|)$  in (26)  is an asymptotic series
$\sum_{k\geq 0} a_k(v,\xi_2)|\eta|^{-k}$,
where
$$a_0(v,\xi_2)=
{(-1)^3\over 8\pi^2}\chi_0(z(v,\xi_1),\hat\eta)e^{i[\alpha(L)+{\pi\over 4}\sigma_-
-{\pi\over 4}]}
\big|\hbox{det}{\partial(v,\xi_2)\over \partial z}\big|^{-{1\over 2}}.\eqno(27)
$$
Note that the factor $e^{-i{\pi\over 4}}$  arises because
$S_{\xi_2^2}(v,p^-(v,w),L)>0$  (cf.  (66.44)  in [E]).
\vskip.1in
To evaluate the contribution of $\int_{\Omega_R}V_0e^{-x\cdot\eta}dvdw$
we apply the stationary phase method to the double integral in
$\xi_2$ and $w$.  The phase function is $S(v,\xi_2,t)+w\xi_2-v$.  The equations
for  the stationary points are
$$
S_{\xi_2}(v,\xi_2,t)+w=0,\ \ \xi_2=0.
$$
Note that  $t=L$.  We will show $w=-S_{\xi_2}(v,0,L)={R\over 2}$:
Let
$\xi_2-\alpha(v)=0$ be the equation of the caustic set,  i.e.
$S_{\xi_2^2}(v,\alpha(v),L)=0$.  In our situation $S_{\xi_2^3}(v,\alpha(v),L)\neq 0$.
Expand $S_{\xi_2}(v,\xi_2,L)$  by the Taylor's formula with a remainder  at $\xi_2=\alpha(v)$.
When $\xi_2=0$, that gives $S_{\xi_2}(v,0,L)=S_{\xi_2}(v,\alpha(v),L)+ c(v)(0-\alpha(v))^2$.
Therefore
$S_{\xi_2}(v,\alpha(v),L)=S_{\xi_2}(v,0,L)- c(v)\alpha^2(v)$.
The equation of the caustic set in $(v,w)$  coordinates is
$w=-S_{\xi_2}(v,\alpha(v),L)=-S_{\xi_2}(v,0,L)+ c(v)\alpha^2(v)$.
On the other hand, using the mapping $x(\tilde v,\tilde w)$, one sees that near $(v,w)=(v_0,R/2)$ with $v_0=-R/(4\sqrt 3)$,
 the caustic set $\Sigma$ is given by
$$
w={R\over 2}-c_1(v)(v-v_0)^2.
$$
Comparing these two expressions for the caustic set we get
$-S_{\xi_2}(v,0,L)={R\over 2}$  and $\alpha(v)=c_2(v)(v-v_0)^2$.  Note that the determinant
of the Hessian at the critical point $(0,{R\over 2})$  is $-1$.
Therefore the standard stationary phase lemma in $(\xi_2,w)$  gives the
asymptotic expansion $\sum_{i\geq 0}r_k^0(v)|\xi|^{-{1\over 2}-k}$,  where
$$
r_0^0={(-1)^3\over 8\pi^2}\big({2\pi\over |\eta|}\big)^{1\over 2}
\chi_0(v\hat \eta +{R\over 2}\hat\eta^\perp,\eta) e^{i(\alpha(L)+{\pi\over 4}\sigma_-
-{\pi\over 4})}
\big({4\sqrt 3\over R}\big)^{-{1\over 2}}.\eqno(28)
$$
In (28) we substituted the value of the Jacobian  in (27). By (11) that is equal  to ${4\sqrt 3\over R}$ at $\xi_1=0, w={R\over 2}$.
\vskip.1in
Combining the contributions of (28)  for $w={R\over 2}$  and $w=-{R\over 2}$  with
the contribution of (25)  and  then integrating in $(v,\theta)$  we get  the leading
terms of the contribution of $W_+(t)$  to the trace:
$${1\over (2\pi)^2}(R\sqrt 3)(2\pi))(2\pi)^{1/2}({R\over 4\sqrt 3})^{1/2}\int_0^\infty \cos(\alpha(L))e^{(i[(L-t)|\eta|-\pi/4]}|\eta|^{1/2}d |\eta|
 \eqno{(29)}$$
This agrees with (19), and therefore the final form of the singularity is again the one given in (3) -- with $\pm$ replaced by a minus sign.
\vskip.1in
Note that contributions from neighborhoods on reflection points can be treated by
introduction of the natural angular coordinate place of $w$ as in the final part of \S 4.
\vskip.2in
\noindent {\bf \S 6. The Aharonov-Bohm Effect on a Torus.}
\vskip.1in
The Aharonov-Bohm effect only arises when the underlying domain is not simply connected. In the previous sections the domain was an annulus. Here we consider the Schr\"odinger operator on a torus. Let
$L=\{m_1e_1+m_2e_2:\ m\in \Bbb Z^2\}$, where $\{e_1,e_2\}$ is a basis for $\Bbb R^2$. We assume that the lattice $L$ has the property: For $d,d^\prime\in L$, if $|d^\prime|=|d|$, then $d^\prime=\pm d$. This is a generic condition that implies that the group of isometries of $L$ consists of lattice translations and the inversion $d\to -d$. Associated to $L$ one has the dual lattice $L^*=\{\delta\in \Bbb R^2: \delta\cdot d\in \Bbb Z \hbox{ for all } d\in L\}$.

\vskip.1in
We consider the Schr\"odinger operator,
$$H_{A,V}={1\over 2}(i\partial_{x_1}+A_1(x))^2+ {1\over 2}(i\partial_{x_2}+A_2(x))^2-V(x),$$
acting on functions on $\Bbb T^2=\Bbb R^2/L$.
The functions $A=(A_1,A_2)$ and $V$ are assumed to be smooth on $\Bbb T^2$
and hence they have smooth extensions to $\Bbb R^2$ satisfying $A(x+d)=A(x)$
and $V(x+d)=V(x)$ for all $d\in L$. As before we assume that the magnetic field vanishes
$$\partial_{x_2}A_1-\partial_{x_1}A_2=0 \hbox{ on }\Bbb T^2.\eqno (30)$$
Thus for any closed curve $\gamma$ on $\Bbb T^2$ the flux
$$\alpha_\gamma=\int_\gamma A(x)\cdot dx,$$
is determined by the homology class of $\gamma$. We let $\gamma_1$ and $\gamma_2$ be a basis
 for the homology group, for instance $$\gamma_j=\{te_j,t\in [0,1)\},\ j=1,2,\eqno (31)$$ and denote the corresponding fluxes by $\alpha_1$ and $\alpha_2$.

Let $g(x)\in C^\infty(\Bbb T^2)$ be such that $|g(x)|=1$. The conjugation of
$H_{A,V}$ by the unitary operator of multiplication by $g(x)$ transforms $H_{A,V}$ to
$H_{\tilde A,V}$, where
$\tilde A=A+ ig^{-1}\nabla g$.
  The condition
$|g(x)|=1$ on $\Bbb T^2$ implies that $g(x)=\exp (2\pi i\delta\cdot x + \varphi(x))$,
where $\delta\in L^*$ and $\varphi(x)$ is periodic. Hence
$\alpha_1(\tilde A)=\alpha_1(A)-2\pi \delta\cdot e_1,
\alpha_2(\tilde A)=\alpha_2(A)-2\pi\delta\cdot e_2$.
Therefore if $A$ and $\tilde A$ are gauge equivalent we have
$$ \alpha_j(\tilde A)=\alpha_j(A) \ \hbox{ modulo}\ \ 2\pi, \ j=1,2.\eqno (32)$$

Expanding $A(x)$ in a Fourier series we have
$$ A(x)=A_0 +\sum_{\delta\in L^*\setminus \{0\}} A_\delta e^{2\pi i \delta\cdot x}, $$
where
$A_0=|\Bbb T^2|^{-1}\int_{\Bbb T^2}A(x)dx,\ |\Bbb T^2|$ denotes the area of
$\{ se_1+te_2; 0\leq s,t\leq 1\}$. Since
$\partial_{x_2}A_1=\partial_{x_1}A_1$ we have $A(x)=A_0+\nabla\varphi(x)$, where
$$
\varphi(x)=\sum_{\delta\in L^*\setminus\{O\}}{\delta\cdot A_\delta\over
2\pi i\delta\cdot\delta}e^{2\pi i\delta\cdot x}.
$$
Therefore when (30) holds $A(x)$ is gauge equivalent to the constant
potential $A_0$. Two constant magnetic potentials $A_0$ and $\tilde A_0$ are not
gauge equivalent if (32) does not hold.
When $\tilde A_0$ is not gauge equivalent to either $A_0$ or $-A_0$ the potentials
  $A_0$ and $\tilde A_0$ have
a different physical impact, in particular, the spectra of $H_{A_0,V}$
and $H_{\tilde A_0,V}$ are not the same.

The last assertion is a consequence of the following theorem.
\vskip.1in
\noindent {\bf Theorem 5.1.}
Suppose (30) holds.
The spectrum of $H_{A,V}$ as a self-adjoint operator on $L^2(\Bbb T^2)$
determines $\cos\alpha_1$ and $\cos\alpha_2$, where
$\alpha_j=\int_{\gamma_j}A(x)\cdot dx,\ j=1,2.$
\vskip.1in
Theorem 5.1 complements the results of [G], [ER1] and [E1]. In particular it shows that, if $A$ and $\tilde A$ give rise to zero magnetic fields on $\Bbb T^2$
but different values for $\cos\alpha_1$ and $\cos\alpha_2$,
the Schr\"odinger operators, $H_{A,V}$ and $H_{\tilde A,V}$ will have different spectra.
This proves the Aharonov-Bohm effect on the torus.
\vskip.1in
  \noindent {\bf Proof of Theorem 5.1.} As in the preceding sections we start with the wave trace formula
  $$\sum_{j=1}^\infty\cos(t\sqrt{\lambda_j})=\int_{\Bbb T^2}E_{\Bbb T^2}(x,x,t)dx,$$
 where $\{\lambda_j\}_{j=1}^\infty$ is the spectrum of $H_{A,V}$ on $\Bbb T^2$ and $E_{\Bbb T^2}(x,y,t)$ is the solution to
 $E_{tt}+H_{A,V}E=0$ on $\Bbb T^2\times \Bbb R$ satisfying $E(x,y,0)=\delta(x-y)$ and $E_t(x,y,0)=0$. Note that
 $$E_{\Bbb T^2}(x,y,t)=\sum_{d\in L}E_{\Bbb R^2}(x+d,y,t),$$
 where $E_{\Bbb R^2}$ is the solution to
 $E_{tt}+H_{A,V}E=0$ on $\Bbb R^2\times \Bbb R$ satisfying $E(x,y,0)=\delta(x-y)$ and $E_t(x,y,0)=0$ when $H_{A,V}$ has been extended to $\Bbb R^2$ by making its coefficients periodic, i.e. $A(x+d)=A(x)$ and $V(x+d)=V(x)$ for all $d\in L$. Hence
 $$\int_{\Bbb T^2}E_{\Bbb T^2}(x,x,t)dx=\sum_{d\in L}\int_{\Bbb T^2}E_{\Bbb R^2}(x+d,x,t)dx. $$
 Since $E_{\Bbb R^2}$ is smooth off the cone $|x-y|^2=t^2$, and our assumption on $L$ implies that only two lattice vectors can have $|d|^2=t^2$ for a fixed value of $t$, the singularity in the wave trace at $t=|d|$, must come from (cf. [ERT], [ER2])
 $$\int_{\Bbb T^2}E_{\Bbb R^2}(x+d,x,t)dx+\int_{\Bbb T^2}E_{\Bbb R^2}(x-d,x,t)dx.$$

 To compute the leading singularities in this trace we will use the Hadamard-H\"ormander parametrix (cf. [H\"or]). We have
 $$E_{\Bbb R^2}(x,y,t)=\partial_t(E_+(x,y,t)-E_+(x,y,-t)),$$
where $E_+$ is the forward fundamental solution. The Hadamard-H\"ormander para-

  \noindent metrix construction for $E_+$ writes $E_+$ as an asymptotic sum of terms with increasing regularity. The first term is $a_0(x,y)e_0(|x-y|,t)$, where
$$e_0={1\over 2\sqrt \pi}(t^2-|x-y|^2)_+^{-1/2}\hbox{ when }t>0 \hbox{ and }e_0=0 \hbox{ when } t<0, \hbox{ and }
$$ $$a_0(x,y)=\exp(i\int_0^1(x-y)\cdot A(y+s(x-y))ds).$$
Therefore (cf. [ER1]) the singularity of the trace at $t=|d|$ determines $I(d)+I(-d)$ where
$$I(d)=\int_{\Bbb T^2}\exp(i\int_0^1d\cdot A(x+sd)ds)dx.$$


Since $A(x)=A_0+\nabla\varphi(x)$, where $\varphi(x)$ is periodic, we have
$$\int_0^1d\cdot A(x+sd)ds = d\cdot A_0 \ \ \ \hbox{ since } \ \
\int_0^1d\cdot\nabla \varphi(x+sd)ds=0$$.

Therefore
$I(d)=e^{id\cdot A_0}|\Bbb T^2|$ and hence the singularity of the wave trace at
$t=|d|$ determines $\cos (A_0\cdot d)$ for all $d\in L$. In particular,
 when $d=e_j$ and $\gamma_j=\{ te_j,t\in [0,1)\}, j=1,2,$ we get
$\alpha_j=\int_{\gamma_j}A(x)\cdot dx=e_j\cdot A_0$. Thus the singularities
of the wave trace when $t=|e_j|$ determine $\cos\alpha_j$ for
$j=1,2$. When $V(x)=V(-x)$, then $H_{A_0,V}$ and $H_{-A_0,V}$ are isospectral and one can only recover $\cos\alpha_j, \ j=1,2,$ from the spectrum. When $V$ is not even, the question of whether one could recover $\exp(i \alpha_j)$, $j=1,2,$ from the spectrum is open.

\end
\newpage

\Refs
\widestnumber\key{XXXX}



\ref\key AB {}\by Y. Aharonov and D. Bohm \paper Significance of electromagnetic potentials
in quantum theory \jour Phys. Rev. {115} (1959), 485
\endref


\ref\key CRR
\by M. Combescure, J. Ralston, D. Robert
\paper A proof of the Gutzwiller semiclassical trace formula using coherent states decomposition
\jour Comm. Math. Phys. {202} (1992), 463-480
\endref

\ref\key CR
\by M. Combescure, D. Robert
\book Coherent States and Applications in Mathematical Physics, Springer-Verlag, Berlin (2012)
\endref

\ref \key D
\by J.J. Duistermaat
\paper Oscillatory integrals, lagrangian distributions and unfolding of singularities
\jour
CPAM {27} (1974), 207-281
\endref


\ref \key DG
\by J.J. Duistermaat, V. Guillemin
\paper The spectrum of positive elliptic operator and periodic bicharacteristics
\jour
Invent. Math. {29} (1975), 39-79
\endref


\ref \key E
\by G. Eskin,
\book Lectures on Partial Differnetial Equations, AMS, Providence (2011)
\endref


\ref \key E1 \by G. Eskin, Inverse spectral problem for the Schr\"{o}dinger
 equation with periodic vector potential \paper Comm. Math. Phys. { 125}
 (1989), 263-300
\endref

\ref \key E2 \by G. Eskin
\paper A simple proof of magnetic and electric Aharonov-Bohm effect
\jour Comm. Math. Phys. {321} (2013), 747-767
\endref

\ref \key EIO \by G. Eskin, H. Isozaki and S. O'Dell
\paper Gauge equivalence and inverse
scattering for \linebreak Aharonov-Bohm effect
 \jour Comm. in PDE { 35}(2010), 2164-2194
\endref

\ref \key ER1 \by G. Eskin, J. Ralston
\paper Remark on spectral rigidity for magnetic Schr\"odinger operators,
\jour Mark Krein Centenary Conference, vol. 2, 323-329,
Operator Theory Adv. Appl., 191, Birkhouser, Basel, 2009
\endref


\ref{}\key ER2 \by G. Eskin, J. Ralston \paper Inverse spectral problems
in rectangular domains \jour Comm. in PDE { 32}(2007), 971-1000
\endref

\ref\key ERT \by G. Eskin, J. Ralston, E. Trubowitz
\paper
On isospectral periodic potentials in $\Bbb R^n$, I and II
\jour
Com. Pure and Appl. Math. {\bf 37}(1984),
647-676,715-753
\endref






\ref \key G \by V. Guillemin \paper Inverse spectral results on
two-dimensional tori \jour Journal of the AMS, { 3}(1990),375-387
\endref


\ref \key GM
\by V. Guillemin, R. Melrose
\paper The Poisson summation formula for manifolds with boundary
\jour Advances in Math., 32(1979), 204-232
\endref


\ref \key H \by B. Helffer \paper Effet d'Aharonov-Bohm sur un \'etat born\'e de
l'equation de Schr\"odinger
\jour
Comm. Math. Phys. 119(1988), 315-329
\endref

\ref \key H\"or \by L. H\"ormander \book The
Analysis of Linear Partial Differential Operators, I-IV,
Springer-Verlag, Berlin (1985)
\endref

\ref \key LC
\by R. Lavine, M. O'Carrol
\paper
Ground state property and lower bounds on energy levels of particle in
a uniform magnetic field and external potential
\jour
J. Math. Phys. { 18} (1977), 1908-1912.
\endref

\ref \key L
\by D. Ludwig,
\paper
Uniform asymptotic expansions at a caustic,
\jour
Comm. Pure Appl. Math.  {10} (1966), 215-266
\endref


\ref \key MF
\by V.P. Maslov, M.V. Fedoriuk
\book Semi-Classical Approximation in Quantum Mechanics, D. Reidel, Dordrecht, (1981)
\endref

\ref \key MS
\by R. Melrose, J. Sj\"ostrand
\paper Singularities of boundary value problems
\jour
I Comm. Pure Appl. Math 31 (1978), 93-617,
II Comm. Pure Appl. Math 35 (1982), 129-168
\endref


\ref \key N \by F. Nicoleau
\paper An inverse scattering problem with the Aharonov-Bohm effect
\jour J. Math. Phys.{41} (2000), 5223-5237
\endref



\ref \key R
\by J. Ralston
\paper Gaussian beams and propagation of singularities,
\book Studies in PDE, MAA Studies in Math. {23} (1982), 207-248
\endref




\ref \key RY \by Ph. Roux and D. Yafaev
\paper On the mathematical theory of the Aharonov-Bohm effect
\jour J. Phys. A: Math. Gen. { 35} (2002), 7481-7492
\endref

\ref \key T
\by A. Tonomura, N. Osakabe, T. Matsuda, T. Kawasaki, J. Endo, S. Yano, and H. Yamada
\paper Evidence for Aharonov-Bohm effect with magnetic field completely shielded
from electron wave
\jour Phys. Rev. Lett., 56, (1986), 792
\endref

\ref \key W
\by R. Weder
\paper The Aharonov-Bohm effect and time-dependent inverse scattering theory
\jour Inverse Problems, {18} (2002), 1041-1056
\endref


\endRefs
